\documentstyle[11pt,aaspp4]{article}
\slugcomment{To appear in ApJ Letters}
\begin{document}

\title{A Methane Imaging Survey for T Dwarf Candidates in $\rho$ Ophiuchi}

\author{Karl E. Haisch Jr.\altaffilmark{1}}
\affil{Utah Valley University, Physics Dept., 800 W. University Pkwy., Orem, Utah  85058-5999, Karl.Haisch@uvu.edu}

\and

\author{Mary Barsony\altaffilmark{1,2}}
\affil{San Francisco State University, Dept. of Physics \& Astronomy, 1600 Holloway Ave., San Francisco, California 94132, mbarsony@stars.sfsu.edu}

\and

\author{Chris Tinney\altaffilmark{1}}
\affil{Department of Astrophysics, School of Physics, University of New South Wales, NSW 2052, Australia, c.tinney@unsw.edu.au}

\altaffiltext{1}{Visiting Astronomer at the Anglo-Australian Telescope, Siding Spring, Australia.}

\altaffiltext{2}{Space Science Institute, 4750 Walnut Street, Suite 205, Boulder, CO 80301}

\begin{abstract}

We report the results of the first deep, wide-field, near-infrared methane imaging survey of the  $\rho$ Ophiuchi cloud core to search for T dwarfs. Among the 6587 objects detected, 22 were identified as T dwarf candidates. Brown dwarf models indicate that at the age and distance of the $\rho$ Ophiuchi cloud, these T dwarf candidates have masses between 1 and 2 Jupiter masses. If confirmed as genuine T dwarfs, these objects would be the youngest, lowest mass, and lowest gravity free-floating objects ever directly observed. The existence of these candidates suggests that the initial mass function of the $\rho$ Ophiuchi cloud extends well into the regime of planetary mass objects. A large fraction (59\% $\pm$ 16\%) of our T dwarf candidates appear to be surrounded by circumstellar disks, and thus represent the lowest mass objects yet found to harbor circumstellar disks.

\end{abstract}


\section{Introduction}

One of the most exciting frontiers of astrophysics research today is the discovery and characterization of sub-stellar objects, ranging from brown dwarf to planetary mass objects (planemos). Of particular interest is a class of object known as a T dwarf.  Field T dwarfs are the coolest (500 K $\leq$ T$_{eff}$  $\leq$ 1400 K) and least luminous brown dwarfs that are directly observable (Vrba et al. 2004; Golimowski et al. 2004; Burgasser et al. 2006; Leggett et al. 2009). The cool atmospheres of T dwarfs are rich in molecular gases, especially methane and water vapor, and condensate clouds (Ackerman \& Marley 2001). In fact, strong, broad methane absorption lines in the near-infrared at 1.3 - 1.4, 1.6 - 1.8, and 2.2 - 2.5 microns represent the distinguishing feature of T dwarfs from hotter brown dwarfs, in which the production of atmospheric methane is prohibited by collisional dissociation (Noll et al. 2000). The methane features that {\it define} T dwarfs are so broad and distinctive that the use of dedicated filters to detect them was proposed soon after the discovery of the T dwarf prototype Gliese 229B (Nakajima et al. 1995; Rosenthal et al. 1996).

Differential methane-band imaging (the acquisition of images in both a methane band and a nearby continuum band) provides an efficient method for detection of isolated ultracool dwarfs. Since both bands can be observed nearly simultaneously, at identical airmasses, in similar seeing and transparency conditions, systematic uncertainties in aperture corrections due to seeing variations, extinction corrections due to airmass variations, and (to some extent) zero-point corrections due to transparency variations can all be canceled out. Furthermore, effects of reddening that have plagued near-infrared imaging searches for low-mass objects in the past (e.g., Luhman 2003) are minimized, since the methane and continuum band filters are adjacent to each other within the $H$-band. A detailed discussion of the method of differential methane-band imaging can be found in Tinney et al. (2005).

Most of the field T dwarfs discovered using methane imaging to date have been from follow-up observations of large-area sky surveys (e.g., Tinney et al. 2005 and references therein).  Only a few investigators, so far, have searched for T dwarfs in young (1 - 10 Myr) star clusters (Zapatero Osorio et al. 2002; Mainzer \& McLean 2003; Lucas et al. 2006; Greissl et al. 2007; Burgess et al. 2009; Scholz et al. 2009; Marsh et al. 2010). Surveying young clusters using methane imaging has several advantages. First, low mass brown dwarfs are more luminous, and thus 2 - 5 orders of magnitude brighter, at this stage of their evolution than those in the general field. Thus, they can be detected at greater distances than field objects. In addition, brown dwarfs in clusters have known ages, as opposed to field brown dwarfs that may have age errors of 5 billion years or more. This is a huge advantage since, for a given luminosity, knowledge of an object's age is necessary to assign a unique mass. Young clusters also have smaller physical sizes than older optical clusters, minimizing the effect of contamination by foreground and background field brown dwarfs.

With these advantages in mind, we have initiated the first deep, wide-field, near-infrared (NIR) methane filter imaging search for T dwarf candidates in the youngest (1 - 3 Myr) and nearest (d $<$ 250 pc) star-forming regions to Earth. At this age, these brown dwarfs are at their brightest. According to isochrones for the COND model of Baraffe et al. (2003), they have planetary masses. In this Letter, we present the results of our methane imaging survey of the $\rho$ Ophiuchi cloud core. With a distance of 120 - 130 pc (Loinard et al. 2008; Mamajek 2008; Lombardi et al. 2008), this cloud is the closest star-forming region to Earth that has a compact core harboring several hundred $\sim$ 1 Myr young stellar objects (Barsony et al. 1997; Wilking et al. 1989; Bontemps et al. 2001; Barsony et al. 2005).

%At ages of $\sim$ 1 - 10 Gyr, however, such objects have masses above the planetary mass boundary.

\section{Observations and Data Reduction}

Near-infrared $J, K_{s}$ (1.25, 2.14 $\mu$m), CH$_{4}$s (1.59 $\mu$m), and CH$_{4}$l (1.673 $\mu$m) observations of the $\rho$ Ophiuchi cloud core were obtained during the period 2008 May 23 - 26 with the IRIS2 NIR imager/spectrograph on the Anglo-Australian Telescope (AAT) 4 m telescope. The CH$_{4}$s and CH$_{4}$l  filters have bandwidths of 0.121 and 0.122 $\mu$m, respectively. IRIS2 consists of a Hawaii HgCdTe 1024$\times$1024 array which, when mounted at the f/8 Cassegrain focus on the AAT, yields a plate scale of 0\farcs45 pixel$^{-1}$ with a corresponding field of view of approximately 7.7$^\prime$$\times$7.7$^\prime$. The FWHM for all observations varied between approximately 2.2 - 3.1 pixels ($\sim$ 1\farcs0 - 1\farcs4). 

Nineteen IRIS2 fields were observed, covering an area of $\sim$ 920 arcmin$^{2}$ on the sky. All fields were observed in the MKO photometric system $J$ and $K_{s}$ filters in a five point dither pattern with 30\arcsec \hspace*{0.05in} offsets between each dither. Integration times at each dither position for the $J$ and $K_{s}$ filters were 15 seconds $\times$ 4 coadds and 6 seconds $\times$ 10 coadds, respectively, for a total integration time in both filters of 5 minutes. Each field was also observed in the CH$_{4}$s and CH$_{4}$l filters in a pseudo-random sixteen point dither pattern within a dither box 30\arcsec \hspace*{0.05in}on a side. The observations were interleaved at each dither position to minimize the number of filter changes. Thus, observations of each field were taken as follows: CH$_{4}$s, CH$_{4}$l, dither, CH$_{4}$l, CH$_{4}$s, dither, and so forth. Integration times at each dither position for both filters were 15 seconds $\times$ 4 coadds.  The entire dither pattern was repeated twice in both filters for a total integration time of 16 minutes in each of the CH$_{4}$s and CH$_{4}$l filters. $H$-band images of each field were constructed by adding the corresponding CH$_{4}$ and CH$_{4}$l images for the given field.

All data were reduced using the Image Reduction and Analysis Facility (IRAF)\footnote[3]{IRAF is distributed by the National Optical Astronomy Observatories, which are operated by the Association of Universities for Research in Astronomy, Inc., under cooperative agreement with the National Science Foundation.}. An average dark frame was constructed from the dark frames taken at the beginning and end of each night's observations. This dark frame was subtracted from all target observations to yield dark subtracted images. Sky frames in each filter were individually made for each observation by median-combining all five $J$ and $K_{s}$ band frames for each field, and the nearest nine CH$_{4}$s and CH$_{4}$l frames in time to the target observation. The individual sky frames were normalized to produce flat fields for each target frame. All target frames were processed by subtracting the appropriate sky frames and dividing by the flat fields. Target frames were then registered and combined to produce the final reduced images in each filter. 

\section{Analysis}

Infrared sources were identified at $K_{s}$-band using the DAOFIND routine within IRAF (Stetson 1987). DAOFIND was run on each field using a FWHM of 2.8 pixels, and a single pixel finding threshold equal to 3 times the mean noise of each image. Each field was individually inspected, and the DAOFIND coordinate files were edited to remove bad pixels and any objects misidentified as stars, as well as to add any missed stars to the list. Objects within 30\arcsec \hspace*{0.05in}of the field edges were also removed from the list, as they were in low signal to noise regions of the image as a result of the dither pattern used. Aperture photometry was then performed on all fields in each filter using the PHOT routine within IRAF. An aperture of 4 pixels in radius was used for all target photometry, and a 10 pixel radius was used for the standard star photometry. Sky values around each source were determined from the mode of intensities in an annulus with inner and outer radii of 10 and 20 pixels, respectively. Our choice of aperture size for our target photometry insured that the individual source fluxes were not contaminated by the flux from neighboring stars; however, they are not large enough to include all the flux from a given source. In order to account for this missing flux, aperture corrections were determined using the MKAPFILE routine within IRAF. The instrumental magnitudes for all sources were corrected to account for the missing flux.

Photometric calibration was accomplished using the list of standard stars of Persson et al. (1998). The standards were observed on the same nights and through the same range of air masses as the $\rho$ Ophiuchi cloud. Zero points and extinction coefficients were established for each night. All magnitudes and colors were transformed to the CIT system using MKO to 2MASS and 2MASS to CIT photometric color transformation equations\footnote[4]{See http://www.astro.caltech.edu/$\sim$jmc/2mass/v3/transformations/}. Because of the extensive spatial overlapping of the cloud images, a number of sources were observed at least twice. We compared the $JHK$, CH$_{4}$s, and CH$_{4}$l magnitudes of 200 duplicate stars identified in the overlap regions. For all stars brighter than the completeness limit of our survey, the photometry of the duplicate stars agreed to within 0.15 magnitudes.

The completeness limit of our observations was determined by adding artificial stars at random positions to each of the 19 fields in all four filters and counting the number of sources recovered by DAOFIND. Artificial stars were added in twelve separate half-magnitude bins, covering a magnitude range of 16.00 to 22.00, with each bin containing 100 stars. The artificial stars were examined to ensure that they had the same FWHM as the real sources in each image. Aperture photometry was performed on all sources to confirm that the assigned magnitudes of the added sources agreed with those returned by PHOT. All photometry agreed to within 0.10 mag. DAOFIND and PHOT were then run and the number of identified artificial sources within each half-magnitude bin was tallied. This process was repeated 20 times. We estimate that our survey is 90\% complete to $J$ = 20.50, $H$ = 20.00, $K_{s}$ = 18.50, CH$_{4}$s = 19.25, and CH$_{4}$l = 19.25.

%\subsection{Astrometry}
%
%Coordinates for all objects were determined relative to the positions of known objects in the 2MASS\footnote[5]{This publication makes use of data products from the Two-Micron All-Sky Survey, which is a joint project of the University of Massachusetts and the Infrared Processing and Analysis Center/California Institute of Technology, funded by the National Aeronautics and Space Administration and the National Science Foundation} catalog. In particular, plate solutions were done using the 2MASS catalog in conjunction with WCSTools, a package of programs and a library of utility subroutines for setting and using the world coordinate system in the headers of the most common astronomical image formats to relate image pixels to sky coordinates\footnote[6]{http://tdc-www.harvard.edu/wcstools/}. The resulting coordinates of all objects in our survey have typical rms uncertainties of $\sim$ 0\farcs5 relative to the coordinates of previously known stars used in their determinations. 

\section{Results}

T dwarf candidates were initially identified by their having methane colors indicative of a T dwarf (e.g., Table 3 in Tinney et al. 2005) and $H$-band magnitudes brighter than the completeness limit of $H$ = 20.0. Models for T dwarfs (i.e., the COND models of Baraffe et al. 2003) indicate that our survey is sensitive down to T8 spectral types,
even through A$_V=$ 10. The resulting list of 565 candidates was further winnowed by examining each individual candidate to ensure that it was indeed fainter in the CH$_{4}$l image as compared to the CH$_{4}$s image. Our final list of 22 T dwarf candidates is shown in Table 1. In the table, we list an ID number for each candidate in column 1, the RA (J2000) and Dec. (J2000) in columns 2 and 3, the near-infrared magnitudes and colors in columns 4 - 11, visual extinction estimates for each candidate T dwarf in column 12, and an estimated spectral type for each candidate based on the methane colors (CH$_{4}$s - CH$_{4}$l) in column 13. The methane colors were calculated as outlined in Tinney et al. (2005). Estimates for A$_{v}$ were computed by dereddening each object in a $JHK_{s}$ color-color diagram as discussed below. In Figure 1, we present CH$_{4}$s ({\it left}) and CH$_{4}$l ({\it right}) images of the latest T spectral type object in Table 1. North is up and east is to the left, with each image centered on the T dwarf candidate. As one can see, the T dwarf candidate is fainter in the CH$_{4}$l image as compared to the CH$_{4}$s image.

In Figure 2, we present the $JHK_{s}$ color-color diagram for the T dwarf candidates in Table 1. In the diagram, we plot the locus of points corresponding to the unreddened main sequence as a solid line and the locus of positions of giant stars as a dashed line (Bessell \& Brett 1988). The two leftmost parallel dashed lines define the reddening band for main sequence stars and are parallel to the reddening vector. Crosses are placed along these lines at intervals corresponding to 5 magnitudes of visual extinction. The classical T Tauri star locus is plotted as a dot-dashed line (Meyer, Calvet, \& Hillenbrand 1997). The reddening law of Cohen et al. (1981), derived in the CIT system and having a slope of 1.692, has been adopted. A significant fraction (13/22; 59\% $\pm$ 16\%) of the T dwarf candidates fall outside and to the right of the reddening lines in the infrared excess region of the diagram. 

\section{Discussion}

We have completed the first deep, wide-field, NIR methane-band imaging survey of the $\rho$ Ophiuchi cloud core to search for candidate planetary mass objects
(planemos). Among the 6587 objects surveyed, 22 were identified as T dwarf candidates. Models for T dwarfs (i.e., the COND models of Baraffe et al. (2003)) indicate that at the age and distance of the $\rho$ Ophiuchi cloud, these T dwarf candidates have masses between 1 and 2 M$_{Jup}$, well into the overlap region between brown dwarfs and extrasolar planets. If confirmed as genuine T dwarfs, these objects would represent the youngest, lowest gravity T dwarfs ever directly observed.

Marsh et al. (2010) have recently presented spectroscopic observations of seven objects in $\rho$ Ophiuchi whose NIR colors were suggestive of low-mass brown dwarfs. One of the seven objects displayed weak, broad absorption features suggestive of methane, and thus an early T spectral type. The COND models suggest a slightly higher mass (2 - 3 M$_{Jup}$) for this source than for our T dwarf candidates. We did not detect this source in the methane filters, and it is therefore not listed in Table 1. The other six sources were detected in our survey, however they did not display methane colors indicative of T dwarfs, consistent with their exclusion as T dwarfs by Marsh et al. (2010).

Our candidate T dwarfs are likely members of the $\rho$ Ophiuchi cloud. Foreground T dwarfs will be much older and less luminous, thus they would only be detectable to distances of $\sim$ 50 pc, making the chances of finding even one very remote. We can also exclude the possibility of foreground contamination given the substantial visual extinction of the candidate objects. In addition, contamination by background objects is minimized in cluster methane imaging studies. The surface density of T dwarfs is $\sim$ one per 100 deg$^{2}$ from SDSS (Knapp et al. 2004). The space density of T dwarfs in the Solar neighborhood inferred from cross-correlating the 2MASS and SDSS surveys is 0.007 pc$^{-3}$, or about one per 140 pc$^{3}$ (Metchev et al. 2008). Our survey region is significantly smaller than this, thereby minimizing foreground/background contamination.

One of the goals of our cluster methane imaging survey is to examine the initial mass function (IMF) in regions that vary in density, from hosting tens of known members in the mass range 1.0 - 0.1 M$_{\odot}$, to hundreds of members. These are numbers which make methane imaging T dwarf surveys both eminently practical and sensitive to the mass function in the range in which the lines between planets and brown dwarfs are so blurred. Determining the relative normalizations between the two mass functions is essentially impossible; however, we can address the {\em shape} of the mass function in this boundary region. That is, we can investigate the question, ``Does the star formation mass function turn over between 40 M$_{J}$ and 5 M$_{J}$"? If it does, then we must determine where it does, as this is critical information needed to refine models of the star formation process. The T dwarf candidates we detected in $\rho$ Ophiuchi appear to suggest that the IMF of this cloud extends well into the regime of planemos.
This finding lends support to the possibility that planetary mass objects can
form via the same process as that for stars, namely, through cloud collapse and
fragmentation.

The hunt for the lowest mass object that can collapse directly from the 
present-day interstellar medium is on ({\it e.g.,} Scholz et al. 2009; Burgess et al. 2009).
On theoretical grounds, there is some critical mass and density below which
cloud fragmentation and gravitational collapse is not feasible ({\it e.g.,} Cruz 2008
and references therein). In this context, it is highly suggestive that none
of our 22 planemo candidates (and none of 9 more planemo candidates
in the CrA cloud identified by us using the same methods) have spectral types later than T6, despite the fact that our surveys are sensitive to spectral types as late as T8.
%,  In addition, the T dwarf candidates we detected are predominantly mid-T spectral types, even though our survey is sensitive to T8 objects. If our candidate sources are confirmed as actual T dwarfs, this would appear to indicate a sharp cut-off at the low-mass end of the IMF of the $\rho$ Ophiuchi star-forming region. This would have important implications for the nature of the IMF at the low-mass end, as it has been suggested from Jean's Mass arguments that cloud fragmentation and collapse is not feasible for fragments below some critical mass and density (see e.g., Cruz 2008, and references therein). 

A large fraction (59\% $\pm$ 16\%) of our T dwarf candidates lie in the infrared excess region of the $JHK_{s}$ color-color diagram. Predictions from both observations and modeling suggest that this is what one would expect from excess emission from circumstellar disks (e.g., Lada \& Adams 1992; Meyer et al. 1997; Haisch et al. 2000). Scholz \& Jayawardhana (2008) have recently observed excess infrared emission around four planetary mass objects in the $\sigma$ Orionis cluster. The masses of these objects range from 8 - 20 M$_{Jup}$, significantly higher than the masses of the candidate T dwarfs in our sample (1 - 2 M$_{Jup}$) which show disk emission. If confirmed as bona fide T dwarfs, our sample objects represent the lowest mass sources yet found to harbor circumstellar disks.

Spectroscopic follow-up of the T dwarf candidates identified through our methane imaging is important, and will be vigorously pursued. NIR spectroscopy will allow us to assign an unambiguous spectral type to each candidate, accurate to a sub-class. In principle, this could be accomplished with methane filter differential photometry alone. In practice, our candidates are close to our detection limits (unlike the case for bright field T dwarfs), where photometric errors are largest, complicating spectral type determination via differential photometry. NIR spectra will also allow us to eliminate extinction effects that skew the methane filter spectral typing. Furthermore, high quality NIR spectra will allow us to constrain the effective temperature of our T dwarf candidates much better than using methane filters alone, and any discrepancies in model fits which might indicate unexpected condensates or perhaps local carbon abundance anomalies will be identified. When complete, our combined methane imaging and spectroscopic survey of our sample young, nearby clusters will probe low-mass star formation in a range of environments and enable robust conclusions to be drawn about the presence or absence of a minimum mass for star formation, and the importance or insignificance of dynamical evolution for young cluster mass functions.

\acknowledgements

We thank the referee for providing helpful suggestions that improved the manuscript. We thank the AAO staff for their outstanding support in making our observations possible. We also thank Utah Valley University undergraduate physics student Sherene Higley for her assistance in the reduction of the AAO data. M. B. gratefully acknowledges NSF grant AST-0206146 which made her contributions to this work possible. Additional support for this work was provided by the National Aeronautics and Space Administration through Chandra Award Number AR1-2005A and AR1-2005B issued by the Chandra X-Ray Observatory Center, which is operated by the Smithsonian Astrophysical Observatory for and on behalf of NASA under Contract NAS8-39073.
%\newpage

\clearpage

\begin{figure}
\plotone{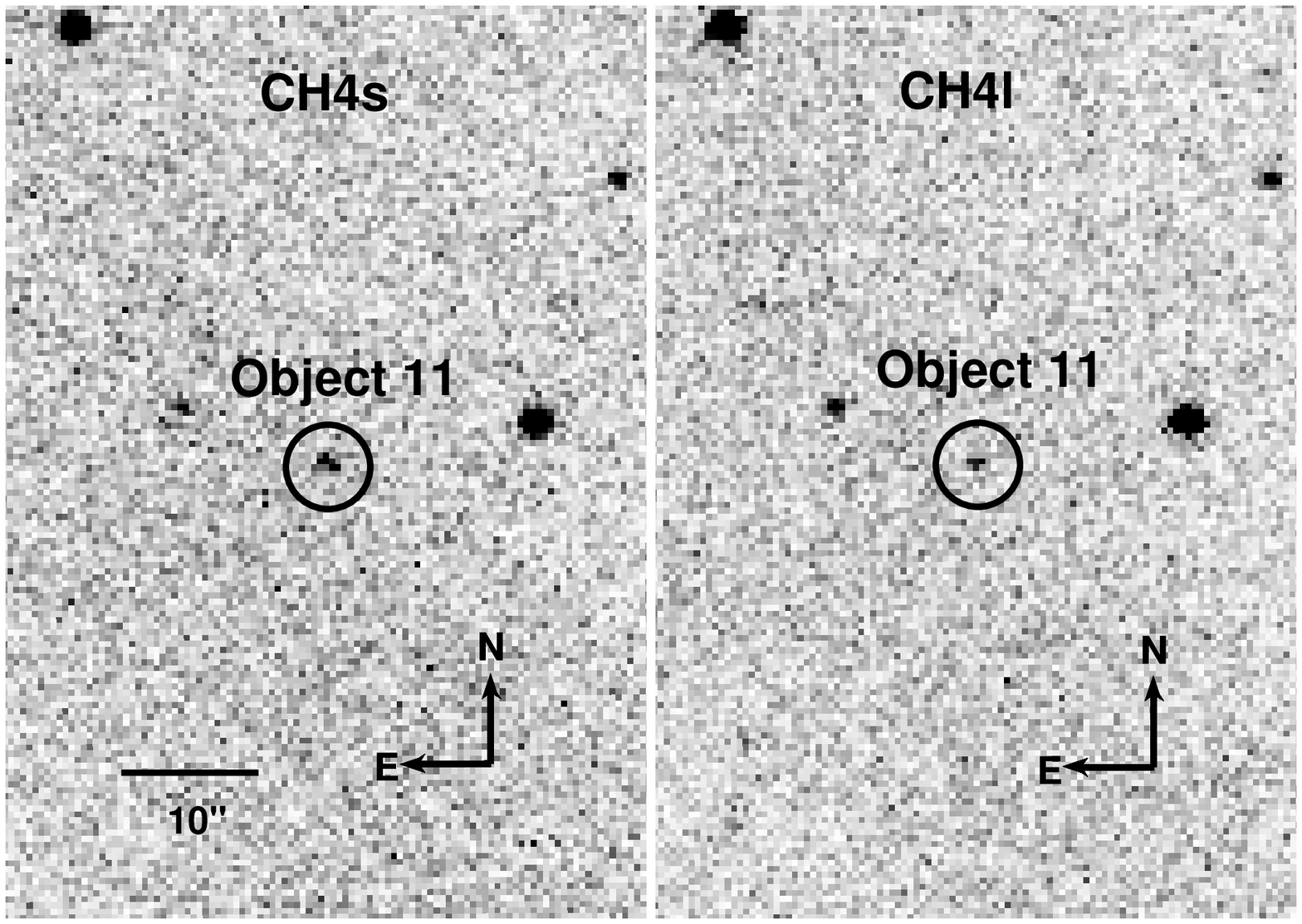}
\caption{CH$_{4}$s ({\it left}) and CH$_{4}$l ({\it right}) images of object 11 from Table 1. North is up and east is to the left in each image. Each image is to the same scale, as indicated by the scale bar in the CH4s image. Object 11 is circled and labelled in each filter. Although it appears asymmetric in the images, the object has a PSF which is Gaussian with a FWHM of 1\farcs1.
Note that this candidate methane absorbing young object is brighter in the CH4s filter than in the CH4l filter, resulting in a negative
CH4s-CH4l color. This behavior is distinctive. The faint background star just to the East of Object 11 is brighter in the CH4l filter than the CH4s filter, as expected for main-sequence dwarfs (an M8 star will have an unreddened positive CH4s-CH4l color of $+$0.17 mags, for example).\label{figure1}}
\end{figure}

\clearpage

\begin{figure}
\plotone{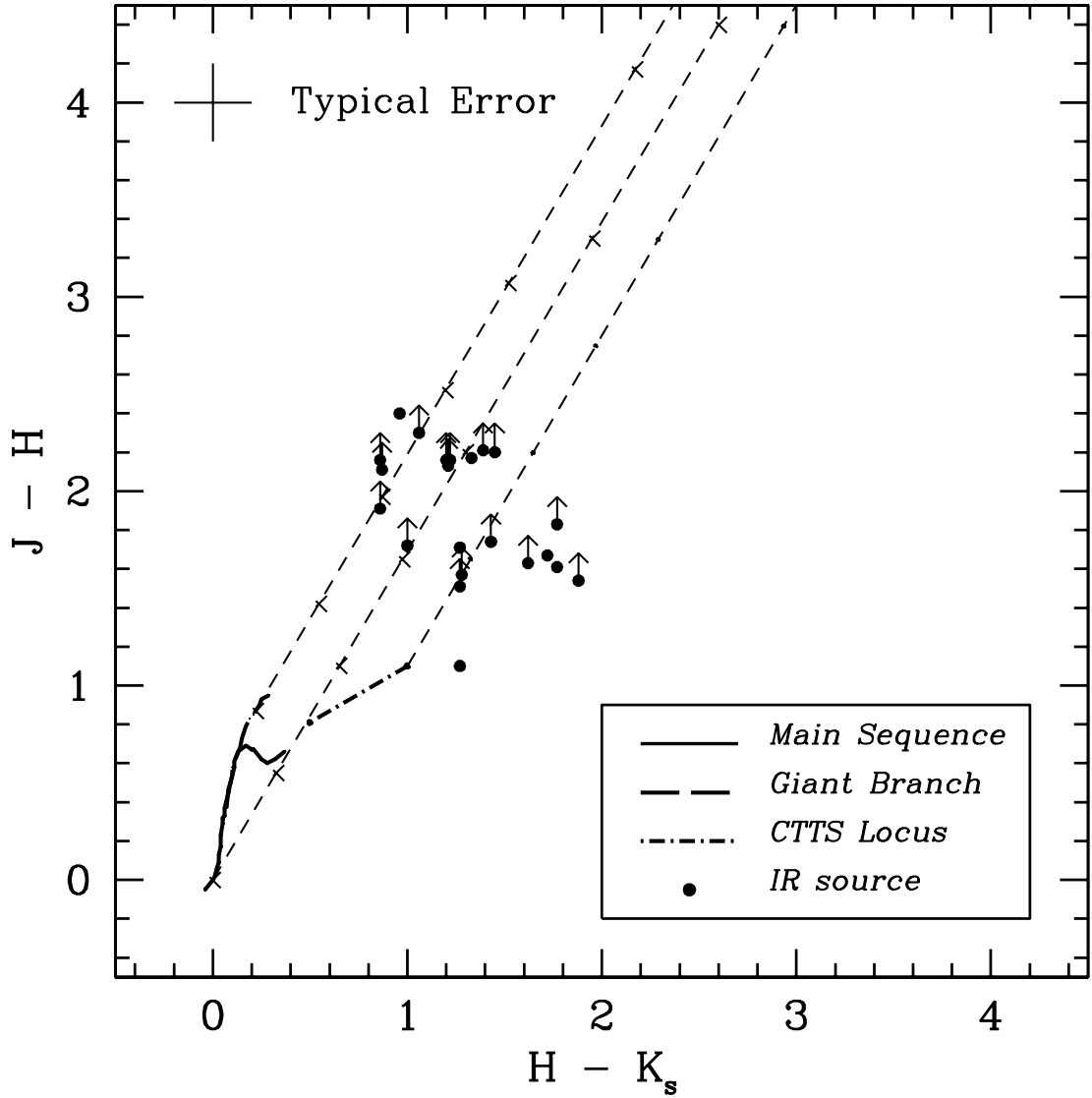}
\caption{$JHK$ color-color diagram for the T dwarf candidates in Table 1. In the diagram, the locus of points corresponding to the unreddened main sequence is plotted as a solid line, the locus of positions of giant stars is shown as a heavy dashed line, and the CTTS locus as a dot-dashed line. The two leftmost dashed lines define the reddening band for main-sequence stars and are parallel to the reddening vector. Crosses are placed along these lines at intervals corresponding to 5 magnitudes of visual extinction. The rightmost dashed line is parallel to the reddening band.\label{figure2}}
\end{figure}

\clearpage

\begin{deluxetable}{rccrccccrccrc}
%\tabletypesize{\scriptsize}
%\rotate
\scriptsize
\tablecaption{Positions, $JHK$, CH$_{4}$s, CH$_{4}$l Magnitudes, and Colors for T Dwarf Candidates\label{table1}}
\tablewidth{0pt}
\tablehead{\colhead{ID} & \colhead{RA (J2000)\tablenotemark{a}} & \colhead{Dec (J2000)\tablenotemark{a}} & \colhead{$J$\tablenotemark{b}} & \colhead{$H$\tablenotemark{b}} & \colhead{$K$\tablenotemark{b}} & \colhead{CH$_{4}$s\tablenotemark{b}} & \colhead{CH$_{4}$l\tablenotemark{b}} & \colhead{$J - H$} & \colhead{$H - K$} & \colhead{CH$_{4}$(s - l)} & \colhead{A$_{v}$\tablenotemark{c}} & \colhead{Sp. T\tablenotemark{d}}}
\startdata
1 & 16:25:49.48  & -24:20:02.09   & 20.98   & 18.58  & 17.72  & 17.66  & 17.89  &  2.40  &  0.86  & -0.43  & 13.2 & T4.5\\
2 & 16:25:55.56  & -24:24:47.45   & $>$21.00   & 19.43  & 18.15  & 18.37  & 18.86  &  $>$1.57  &  1.28  & $<$-0.56  &  $>$5.6 & $>$T5\\
3 & 16:25:55.57  & -24:26:59.57   & $>$21.00   & 19.28  & 18.28  & 18.23  & 18.61  &  $>$1.72  &  1.00  & $<$-0.57  & $>$12.3 & $>$T5\\
4 & 16:25:56.03  & -24:34:00.94   & 20.12   & 18.41  & 17.14  & 17.61  & 17.96  &  1.71  &  1.27  & -0.44  &  6.9 & T4.5\\
5 & 16:25:59.37  & -24:16:16.28   & $>$21.00   & 18.87  & 17.66  & 17.94  & 18.21  &  $>$2.13  &  1.21  & $<$-0.51  & $>$15.6 & $>$T4.5\\
6 & 16:26:15.24  & -24:34:38.21   & $>$21.00   & 18.84  & 17.64  & 17.91  & 18.12  &  $>$2.16  &  1.20  & $<$-0.45  & $>$15.4 & $>$T4.5\\
7 & 16:26:26.76  & -24:39:29.73   & 20.92   & 18.75  & 17.42  & 17.79  & 18.03  &  2.17  &  1.33  & -0.43  & 12.1 & T4.5\\
8 & 16:27:04.19  & -24:27:06.31   & $>$21.00   & 19.46  & 17.58  & 18.41  & 18.92  &  $>$1.54  &  1.88  & $<$-0.58  &  $>$5.4 & $>$T5\\
9 & 16:27:12.11  & -24:17:13.05   & 20.25   & 19.15  & 17.88  & 18.03  & 18.49  &  1.10  &  1.27  & -0.46  &  0.0 & T4.5\\
10 & 16:27:16.62  & -24:22:33.63   & $>$21.00   & 19.37  & 17.75  & 18.29  & 18.55  &  $>$1.63  &  1.62  & $<$-0.34  &  $>$6.2 & $>$T4\\
11 & 16:27:20.41  & -24:19:14.59   & $>$21.00   & 18.80  & 17.35  & 17.59  & 18.35  &  $>$2.20  &  1.45  & $<$-0.93  & $>$11.4 & $>$T6\\
12 & 16:27:24.81  & -24:38:33.78   & $>$21.00   & 19.17  & 17.40  & 17.94  & 18.50  &  $>$1.83  &  1.77  & $<$-0.67  &  $>$8.0 & $>$T5.5\\
13 & 16:27:30.18  & -24:39:24.66   & 20.55   & 18.94  & 17.17  & 17.94  & 18.61  &  1.61  &  1.77  & -0.75  &  6.0 & T5.5\\
14 & 16:27:33.25  & -24:19:03.66   & 20.82   & 19.15  & 17.43  & 18.04  & 18.69  &  1.67  &  1.72  & -0.74  &  6.6 & T5.5\\
15 & 16:27:33.71  & -24:19:30.43   & $>$21.00   & 18.79  & 17.40  & 17.85  & 18.12  &  $>$2.21  &  1.39  & $<$-0.44  & $>$11.5 & $>$T4.5\\
16 & 16:27:35.82  & -24:21:13.07   & $>$21.00   & 19.09  & 18.23  & 17.95  & 18.46  &  $>$1.91  &  0.86  & $<$-0.66  & $>$10.2 & $>$T5.5\\
17 & 16:27:36.71  & -24:24:40.20   & $>$21.00   & 19.49  & 18.22  & 18.31  & 18.95  &  $>$1.51  &  1.27  & $<$-0.70  &  $>$5.1 & $>$T5.5\\
18 & 16:27:41.75  & -24:16:04.89   & $>$21.00   & 18.70  & 17.64  & 17.66  & 17.93  &  $>$2.30  &  1.06  & $<$-0.47  & $>$13.2 & $>$T4.5\\
19 & 16:27:44.27  & -24:21:01.28   & $>$21.00   & 18.84  & 17.62  & 17.88  & 18.12  &  $>$2.16  &  1.22  & $<$-0.49  & $>$15.7 & $>$T4.5\\
20 & 16:28:10.81  & -24:29:04.68   & $>$21.00   & 18.84  & 17.98  & 17.79  & 18.20  &  $>$2.16  &  0.86  & $<$-0.56  & $>$10.2 & $>$T5\\
21 & 16:28:20.79  & -24:38:25.69   & $>$21.00   & 18.89  & 18.02  & 17.79  & 18.12  &  $>$2.11  &  0.87  & $<$-0.48  &  $>$10.7 & $>$T4.5\\
22 & 16:28:45.89  & -24:34:20.44   & $>$21.00   & 19.26  & 17.83  & 18.24  & 18.50  &  $>$1.74  &  1.43  & $<$-0.36  &  $>$7.2 & $>$T4\\
\enddata
\tablenotetext{a}{Coordinates listed are J2000. Units of right ascension are hours, minutes, and seconds, and units of declination are degrees, arcminutes, and arcseconds.}
\tablenotetext{b}{Typical uncertainties in the listed magnitudes are 0.10 - 0.15 magnitudes.}
\tablenotetext{c}{Extinction estimates were calculated by dereddening each source in the $JHK$ color-color diagram as discussed in the text.}
\tablenotetext{d}{Estimated spectral type based on the CH$_{4}$(s - l) colors.}
\end{deluxetable}

\end{document}